\documentclass[aps,preview,showpacs,preprintnumbers,amsmath,amssymb,superscriptaddress,nofootinbib,longbibliography]{revtex4-1}

\usepackage{graphicx}% Include figure files

\begin{document}
\title{Nanoscale X\hbox{-}ray investigation of magnetic metallofullerene peapods}
\author{Fabian Fritz}
	\affiliation{Department of Physics, University Osnabr\"uck, 49076 Osnabr\"uck, Germany}
	\affiliation{Peter Gr\"unberg Institute (PGI-6), Forschungszentrum J\"ulich GmbH, 52425 J\"ulich, Germany}
	\affiliation{JARA - Fundamentals of Future Information Technologies, Forschungszentrum J\"ulich GmbH, 52425 J\"ulich, Germany}
\author{Rasmus Westerstr\"om}
	\affiliation{The division of synchrotron radiation research, Lund University, 22100 Lund, Sweden}
\author{Aram Kostanyan}
	\affiliation{Physik-Institut, University of Z\"urich, CH-8057 Z\"urich, Switzerland}
	\affiliation{Swiss Light Source, Paul Scherrer Institute, 5232 Villigen PSI, Switzerland}
\author{Christin Schlesier}
	\affiliation{Leibniz Institute for Solid State and Materials Research (IFW), 01069 Dresden, Germany}
\author{Jan Dreiser}
	\affiliation{Swiss Light Source, Paul Scherrer Institute, 5232 Villigen PSI, Switzerland}
\author{Benjamin Watts}
	\affiliation{Swiss Light Source, Paul Scherrer Institute, 5232 Villigen PSI, Switzerland}
\author{Lothar Houben}
	\affiliation{Ernst Ruska-Centre for Microscopy and Spectroscopy with Electrons, Forschungszentrum J\"ulich GmbH, 52425 J\"ulich, Germany}
	\affiliation{JARA - Fundamentals of Future Information Technologies, Forschungszentrum J\"ulich GmbH, 52425 J\"ulich, Germany}
\author{Martina Luysberg}
	\affiliation{Ernst Ruska-Centre for Microscopy and Spectroscopy with Electrons, Forschungszentrum J\"ulich GmbH, 52425 J\"ulich, Germany}
	\affiliation{JARA - Fundamentals of Future Information Technologies, Forschungszentrum J\"ulich GmbH, 52425 J\"ulich, Germany}
\author{Stanislav M Avdoshenko}
	\affiliation{Leibniz Institute for Solid State and Materials Research (IFW), 01069 Dresden, Germany}
\author{Alexey A Popov}
	\affiliation{Leibniz Institute for Solid State and Materials Research (IFW), 01069 Dresden, Germany}
\author{Claus M Schneider}
	\affiliation{Peter Gr\"unberg Institute (PGI-6), Forschungszentrum J\"ulich GmbH, 52425 J\"ulich, Germany}
	\affiliation{JARA - Fundamentals of Future Information Technologies, Forschungszentrum J\"ulich GmbH, 52425 J\"ulich, Germany}
\author{Carola Meyer}
	\affiliation{Department of Physics, University Osnabr\"uck, 49076 Osnabr\"uck, Germany}
	\affiliation{Peter Gr\"unberg Institute (PGI-6), Forschungszentrum J\"ulich GmbH, 52425 J\"ulich, Germany}
	\affiliation{JARA - Fundamentals of Future Information Technologies, Forschungszentrum J\"ulich GmbH, 52425 J\"ulich, Germany}
%\ead{carola.meyer@uos.de}
%\ead{rasmus.westerstrom@sljus.lu.se}
%\vspace{10pt}
\begin{abstract}
Endohedral lanthanide ions packed inside carbon nanotubes (CNTs) in a one-dimensional assembly have been studied with a combination of high resolution transmission electron microscopy (HRTEM), scanning transmission X\hbox{-}ray microscopy (STXM), and X\hbox{-}ray magnetic circular dichroism (XMCD). By correlating HRTEM and STXM images we show that structures down to 30\,nm are resolved with chemical contrast and record X\hbox{-}ray absorption spectra from endohedral lanthanide ions embedded in individual nanoscale CNT bundles. XMCD measurements of an Er$_3$N@C$_{80}$ bulk sample and a macroscopic assembly of filled CNTs indicates that the magnetic properties of the endohedral Er$^{3+}$ ions are unchanged when encapsulated in CNTs. This study demonstrates the feasibility of local magnetic X\hbox{-}ray characterization of low concentrations of lanthanide ions embedded in molecular nanostructures.
\end{abstract}
\date{\today}
%\noindent{\it Keywords}: CNT, Endohedral Fullerenes, HRTEM, STXM, XMCD
%\clearpage
\maketitle

\section{Introduction}
Single-molecule magnets (SMMs) are of interest in a variety of research fields, including molecular spintronics \cite{Bogani2008,Urdampilleta2011,Urdampilleta2013} and quantum computing \cite{Thiele2014,Leuenberger2001,Affronte2007,Ardavan2007,Bertaina2008}. However, the development of such technologies requires techniques for mediating the interaction between the SMMs, as well as with the macroscopic world. Carbon nanotubes (CNTs) present a possible solution to this problem. Due to their extreme aspect ratio they have a nanoscale diameter coupled with a micro- to macroscale length, as well as having configurable electronic properties that are sensitive to the local chemical environment. One realization of this idea involves grafting an SMM, such as TbPc$_2$ \cite{Ishikawa2003}, onto the outside wall of a CNT in order to influence the transport properties of the CNT, thereby providing the means for an electronic readout of the spin state \cite{Bogani2008} and creating a supramolecular spin valve \cite{Urdampilleta2011} that allowed for the readout of a single Tb electronic and nuclear spin \cite{Urdampilleta2013}.
\par
Instead of grafting the molecules to the outside walls, SMMs can also be filled into CNTs. This approach protects the molecules from the environment and avoids strong decoherence, which would disturb quantum computing operations \cite{Benjamin2006}. In the first successful experiment, Mn$_{12}$ clusters were filled into large diameter CNTs \cite{Gimenez-Lopez2011}. But due to the size ratio between the SMMs and the CNTs, the molecules appear in an amorphous phase inside the CNTs. Endohedral metallofullerene SMMs, on the other hand, have diameters that closely match CNTs and so packing them into a CNT in a so-called ``peapod'' structure results in a one-dimensional alignment of SMMs. This enables oriented and therefore well-defined coupling between the SMMs similar to the coupling of paramagnetic spins in a chain \cite{Harneit2002a,Harneit2002b}. Rare-earth-based endofullerenes constitute a new class of SMMs \cite{Westerstroem2012,Westerstroem2014,Dreiser2014,Junghans2015,Liu2016}. By varying the cluster stoichiometry \cite{Westerstroem2014}, composition \cite{Junghans2015}, and geometry \cite{Liu2016}, their magnetic properties can be tailored without significantly changing the size of the encapsulating carbon cage, thus providing a large selection of possible SMMs that could be encapsulated.
\par
Even though CNTs provide one route towards confining low-dimensional arrays of SMMs and possibly allowing electronic readout of their magnetic states, questions remain regarding how the magnetic properties may be modified when the SMMs are transferred from the bulk phase to low-dimensional confinements. In particular, it is of interest to know if the magnetic bistability is affected and if intermolecular interactions can lead to a one-dimensional alignment of the molecular magnetic moments.
\par
To investigate the effect of one-dimensional ordering, one has to determine the atomic scale structure and the magnetic properties of single CNTs filled with endofullerene SMMs. High-resolution transmission electron microscopy (HRTEM) enables the structural characterization and was already demonstrated for trimetal nitride fullerenes inside CNTs. Although Ho$_3$N@C$_{80}$ \cite{Gimenez-Lopez2011a} and Er$_3$N@C$_{80}$ \cite{Khlobystov2004,Sato2007} have been imaged with atomic resolution, a magnetic characterization in HRTEM is hampered by distortions induced by the electron optics.
\par
Given a sufficiently small X\hbox{-}ray beam, X\hbox{-}ray magnetic circular dichroism (XMCD) provides a suitable tool for studying magnetism on the nanoscale. XMCD is element specific, sensitive and provides absolute values of spin and orbital angular moments \cite{Thole1992,Piamonteze2012} as well as the orientation of the endohedral units \cite{Westerstroem2015}. With current X\hbox{-}ray optics, it is possible to focus an X\hbox{-}ray beam down to the nanometre scale. Even though a single CNT with dimensions suitable for peapod structures is about one order of magnitude smaller than the lateral resolution of current instruments, assemblies (bundles) of CNTs are observable with scanning transmission X\hbox{-}ray microscopes (STXM).
\par
Previous combined HRTEM and STXM studies of nano carbon systems have mostly been focused on empty CNTs \cite{Felten2007,Liu2014,Zhao2015}, but also CNTs filled with Fe particles \cite{Chen2015}, or with catalyst nanoparticles at the tip or along the CNTs \cite{Gao2014} have been investigated.
\par
Here, we are pushing the resolution and sensitivity limits of STXM by studying low concentrations of magnetic lanthanide ions embedded in one-dimensional peapod structures. We demonstrate that single CNT bundles can be located and studied by HRTEM and STXM \cite{Ade1992}, enabling us to correlate STXM (resolution $<$ 30\,nm) and HRTEM data from the same nanoscale object. The combination of the two local techniques provides images of individual CNTs and endofullerene cages, X\hbox{-}ray absorption (XAS) spectra from single CNT bundles, and a spatial map of the lanthanide content.
\par
Since the current STXM beamline setup does not allow for measurements under the conditions needed to observe SMM properties, and due to the low yield of endofullerene SMMs, this initial feasibility study uses the paramagnetic Er$_3$N@C$_{80}$ endofullerenes that are commercially available in high purity and suitable amounts. However, the methodology described in this work can be applied to CNT systems containing other endofullerene SMMs.
\par
As the first step towards a magnetic characterization of one-dimensional arrays of endofullerenes, we have performed low-temperature XMCD measurements of a macroscopic assembly of CNTs filled with Er$_3$N@C$_{80}$ endofullerenes. Using a bulk sample as reference, our results indicate that the magnetic properties are unchanged when encapsulated in CNTs.
\section{Experimental Section}
We use Er$_3$N@C$_{80}$ (SES Research) with a purity of 97\,w\% for our samples. The CNTs are commercial single-walled carbon nanotubes produced by an electric arc (Carbon Solutions, Inc.). The sample includes a statistical mixture of different chiralities, of semiconducting and metallic CNTs, with a diameter distribution of 1.55\,nm~$\pm$~0.1\,nm resulting in a one-dimensional alignment of the fullerene molecules inside the CNTs. The CNT material contains traces of catalysts consisting of iron and yttrium nanoparticles. Possible dipole-dipole interactions are not expected to significantly influence the magnetic ground state of the small fraction of peapods that are in the proximity to the catalyst nanoparticles. The same applies for paramagnetic defects on the CNTs induced by oxidation.
\par
The filling method is adopted from Khlobystov et al. \cite{Khlobystov2004} and uses supercritical CO$_2$ (50$^\circ$C, 150\,bar) as solvent. The pressure is varied between 100 and 150\,bar to increase the yield. Prior to the filling, the CNTs are oxidized in air for 20\,min at 550$^\circ$C, creating defects which enable the fullerenes to enter the CNTs \cite{Meyer2009}. After the process, the filled CNTs are washed with \textit{ortho}-dichlorobenzene and toluene, and are subsequently filtered through a 450\,nm pore PTFE membrane to remove excess fullerenes.
\par
A filling yield of 33\% could be determined by measuring the amount of erbium inside the sample using inductively coupled plasma optical emission spectrometry.
\par
For HRTEM and X\hbox{-}ray measurements, the filled CNTs are sonicated in methanol for 10\,min and deposited onto Cu TEM grids (Quantifoil, holey carbon grids), which leads to bundles of 20~-~100\,nm diameter. The filled CNT bundles are thus suspended over holes with a diameter of 1\,$\mu$m excluding any substrate interaction. We use these grids for direct correlation of STXM and HRTEM measurements of \textit{the same} site.
\par
Er$_3$N@C$_{80}$ molecules are dissolved in toluene and spray coated on an aluminium plate as a bulk reference sample.
\par
HRTEM measurements are carried out using an FEI TECNAI F20 instrument \cite{tecnai} at an acceleration voltage of 120\,kV. To image the erbium ions, the sample was measured with the spherical and chromatic aberration corrected FEI Titan 60-300 Ultimate (``PICO'') \cite{pico} at 80\,kV. All shown images were taken with a slight underfocus to achieve the highest resolution.
\par
The scanning transmission X\hbox{-}ray microscopy measurements were performed at the PolLux beamline \cite{Raabe2008} of the Swiss Light Source. The images were recorded at room temperature by raster scanning the sample across a 15\,nm focused X\hbox{-}ray beam while detecting the total electron yield (TEY) with a channeltron detector \cite{Hub2010} in order to achieve the sensitivity needed to detect the low concentration of Er$^{3+}$ ions in the sample.
\par
The X\hbox{-}ray absorption measurements were carried out at the X\hbox{-}Treme beamline \cite{Piamonteze2012} of the Swiss Light Source. Absorption spectra were acquired by measuring the TEY in the on-the-fly mode \cite{Krempasky2010}. The reference sample and the filled CNTs were measured with different monochromator settings and the photon energy axis in figure~\ref{figure5a_5e}b has been changed accordingly. For the filled CNT sample, the measurement was performed with a beam spot size of 1\,mm$^2$.
\par
The magnetization curves were computed using PHI code \cite{Chilton2013} which employed crystal-field parameters obtained \textit{ab initio} at the CASSCF/SO-RASSI level using the SINGLE ANISO module \cite{Chibotaru2012} implemented in MOLCAS 8.0.3. Geometry parameters of the molecule were those for DFT-optimized $C_3$-symmetric Lu$_3$N@C$_{80}$ as Lu$^{3+}$ has a similar ionic radius to Er$^{3+}$. \textit{Ab initio} calculations were then performed for the ErY$_2$N@C$_{80}$ molecule. The active space of the CASSCF calculations includes eleven active electrons and the seven active orbitals (e.g. CAS (11,7)). All 35~quartet states and 112~doublets were included in the state-averaged CASSCF procedure and were further mixed by spin-orbit coupling in the RASSI procedure. The atomic natural extended relativistic basis set (ANO-RCC) was employed with VDZ-quality of Y and Er.
\section{Results and Discussion}
\subsection{HRTEM analysis of peapods}
\begin{figure}[hbt]
	\centering
	\includegraphics[width=8.0cm]{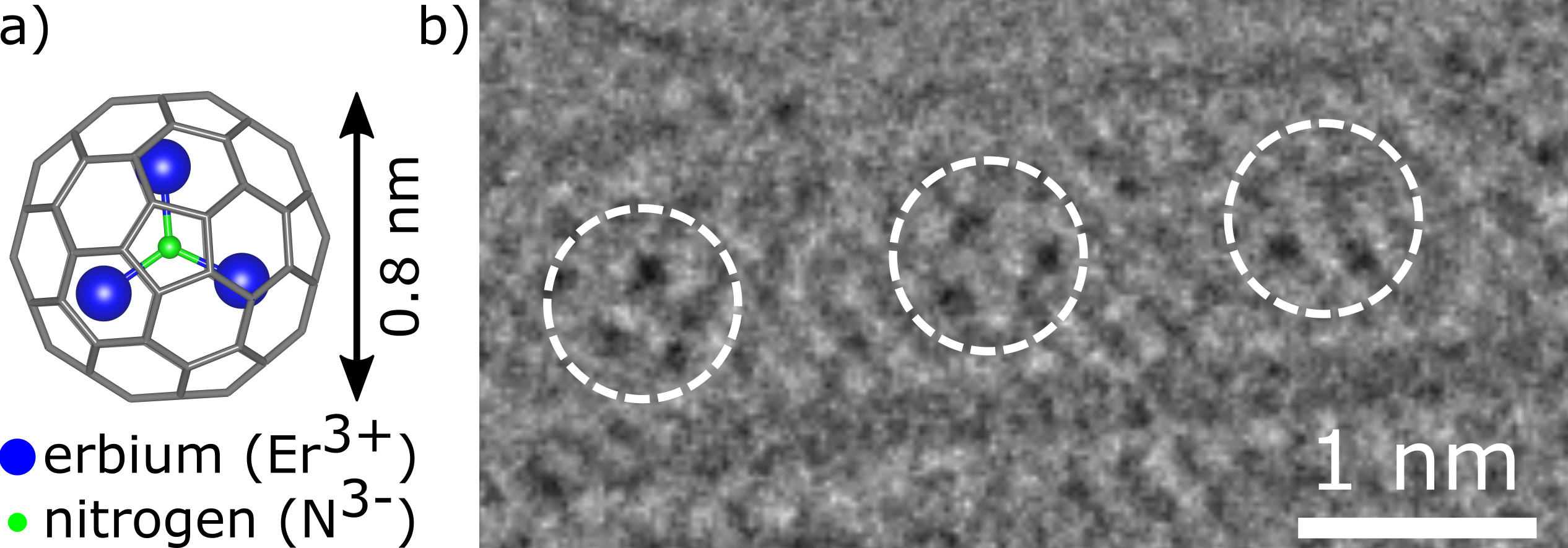}%
	\caption{a) Schematic drawing of one Er$_3$N@C$_{80}$ molecule. b) HRTEM image showing the successful one-dimensional filling of the fullerenes into CNTs. The erbium atoms appear as dark spots due to their high atomic mass. The fullerene cages are marked with white dotted circles for better visibility.}%
	\label{figure1a_1b}
\end{figure}
\noindent
Figure~\ref{figure1a_1b}a shows a schematic drawing of the Er$_3$N@C$_{80}$ molecule. These fullerenes consist of a C$_{80}$ cage filled with a trigonal cluster containing one nitrogen and three erbium atoms. The HRTEM image plotted in figure~\ref{figure1a_1b}b shows the successful filling of the carbon nanotubes. Due to the diameter of the CNTs with respect to the fullerenes, the filling is one-dimensional. The HRTEM also demonstrates that the erbium clusters stay intact inside the CNTs after filling. The distance between the fullerenes in this HRTEM image is a bit larger compared to other examples in the literature \cite{Khlobystov2004,Sato2007}.	This is explained by partial filling of this particular CNT, which enables the fullerenes to move along the tube axis. The HRTEM image is a snapshot of this situation resulting in a larger spacing between the molecules. A much denser filling similar to previous studies is observed in most of the HRTEM images taken on this sample, however the erbium ions are not clearly visible in those images due to rotation of the Er$_3$N cluster.
\subsection{STXM imaging and spectroscopy of isolated CNT bundles}
The STXM measurements were performed using a focused X\hbox{-}ray beam with a dimension of about 15\,nm on the sample. Images were recorded by scanning the sample across the X\hbox{-}ray beam while measuring the total electron yield (TEY) using a channeltron detector. Similarly, X\hbox{-}ray absorption spectra of selected sample areas can be measured by varying the photon energy of the focused X\hbox{-}ray beam. Prior to the X\hbox{-}ray measurements, a pre-characterization of the sample was performed using HRTEM to identify suitable areas to study with STXM.
\par
\begin{figure}[hbt]
	\centering
	\includegraphics[width=8.0cm]{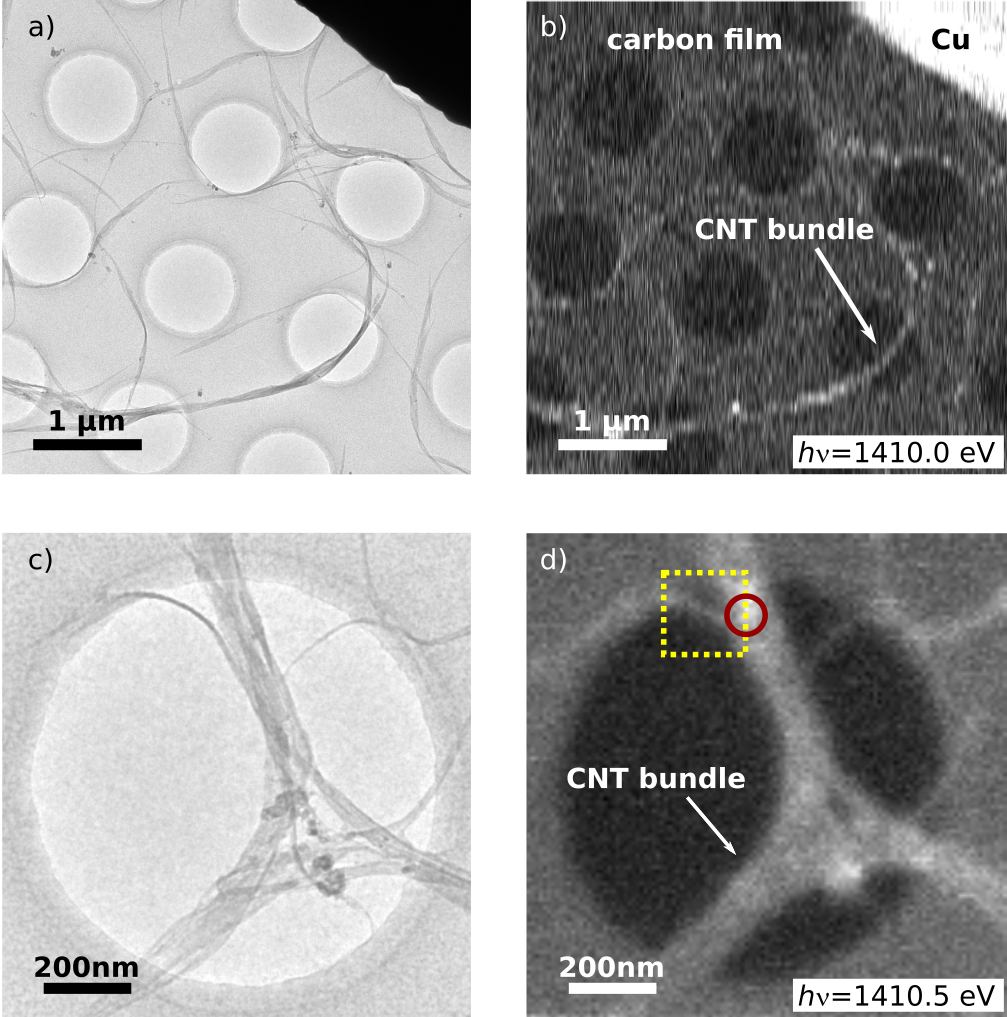}%
	\caption{HRTEM pre-characterization images (a+c) and corresponding STXM images (b+d) of two spots; the pixel size is (10x100)\,nm$^2$ for (b) and (10x10)\,nm$^2$ for (d); the dwell time is 10\,ms and 500\,ms, respectively. At the position of the red circle the XAS spectrum was taken and the yellow square gives the area of the measured erbium map in figure~\ref{figure4a_4d}.}%
	\label{figure2a_2d}
\end{figure}

Figure~\ref{figure2a_2d}a and \ref{figure2a_2d}c show HRTEM images of CNT bundles suspended over the holes in the TEM grid. The corresponding STXM images in figure~\ref{figure2a_2d}b and \ref{figure2a_2d}d clearly reproduce the structural information, demonstrating that bundles down to 20\,nm diameter are visible with the X\hbox{-}ray microscope.
%\clearpage
\par
\begin{figure}[h!]
	\centering
	\includegraphics[width=8.0cm]{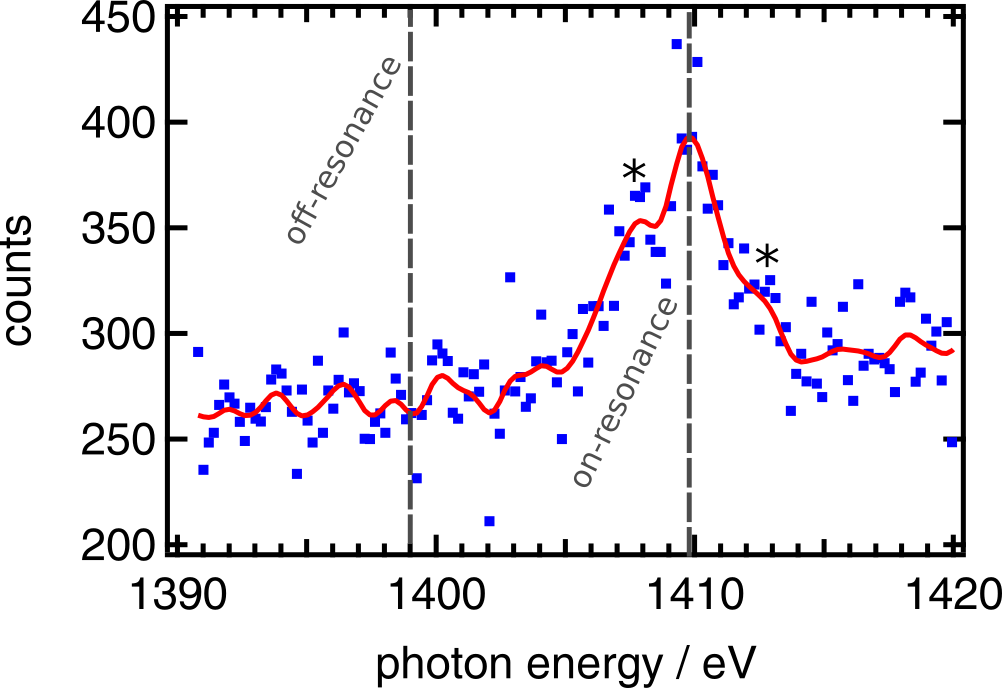}%
	\caption{XAS spectrum from a peapod bundle (marked with a red circle in figure~\ref{figure2a_2d}d) at the Er\,$M_5$\hbox{-}edge. The blue points represent the measured data. The red line is obtained by smoothing the data with Gaussian filtering to clarify the side peaks of the multiplet structure which are marked with asterisks. The marked off- and on-resonance energies are used to record the images in figure~\ref{figure4a_4d}. An energy step size of 0.2\,eV and a dwell time of 1\,s were used.}%
	\label{figure3}
\end{figure}

Figure~\ref{figure3} displays an XAS spectrum from a large CNT bundle at the position marked with a red circle in figure~\ref{figure2a_2d}d. This spectrum represents the first nanoscale X\hbox{-}ray spectrum of an endofullerene peapod structure. A clear peak at the Er\,$M_5$\hbox{-}edge is visible in the spectrum. Despite the low signal-to-noise ratio caused by the low amount of Er probed, the characteristic multiplet structure can still be resolved.
\par
The XAS spectrum demonstrates the existence of erbium at the measured spot, indicating an area on the CNTs with a high filling of Er$_3$N@C$_{80}$. The present XAS spectrum was recorded using linearly polarized light. However, the same experiment could, in principle, be performed using circularly polarized light, demonstrating the feasibility of performing XMCD studies of individual molecular nanostructures.
\par
\begin{figure}[h!]
	\centering
	\includegraphics[width=8.0cm]{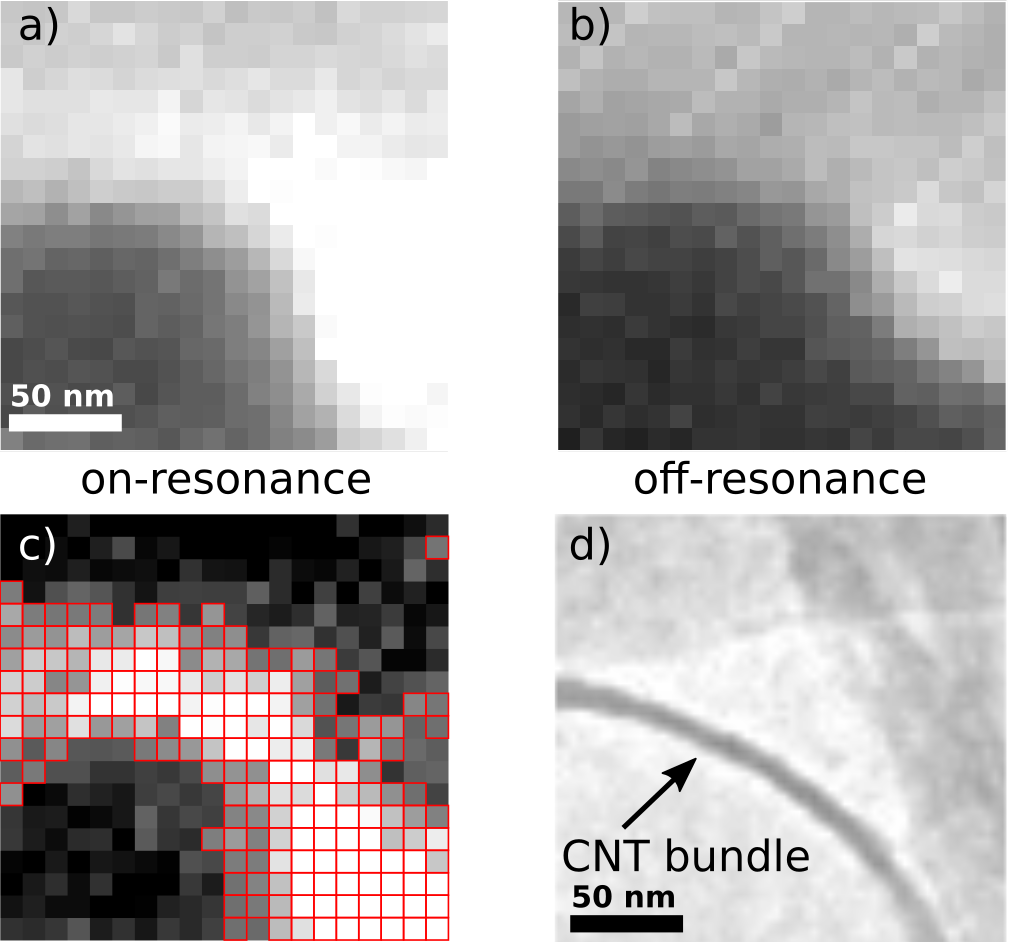}%
	\caption{Map of erbium inside a peapod bundle: (a) and (b) STXM images measured with a pixel size of (10x10)\,nm$^2$ and a dwell time of 20\,s at E~=~1409.7\,eV and E~=~1399.0\,eV, respectively. The colour scale ranges from 6,000~counts (black) to 11,000~counts (white). (c) Difference image [(a)-(b)] giving the positions of erbium at this spot. Grey scale: 500~counts (black) to 2,500~counts (white). The red-framed pixels indicate positions with clear erbium signal (number of counts $>$ 1125). (d) Corresponding HRTEM image of the same spot.}%
	\label{figure4a_4d}
\end{figure}

The STXM technique enables mapping of a specific element in the sample. In our case, a spatial map of erbium ions inside a single peapod bundle is measured (figure~\ref{figure4a_4d}). The chosen area contains a peapod bundle with a diameter of around 30\,nm and is located in the top left corner of the image in figure~\ref{figure2a_2d}c marked with a yellow square.
\par
Figure~\ref{figure4a_4d}a and \ref{figure4a_4d}b show two images recorded with the photon energy set to the Er\,$M_5$\hbox{-}edge (4a) and the pre\hbox{-}edge (4b). The difference between the two images represents a spatial map of the Er content (figure~\ref{figure4a_4d}c). The two recorded STXM images are slightly shifted with respect to each other due to slight drifts during the long measurement time. For calculating the difference, this drift has been corrected. The presence of an erbium signal is indicated by the red-framed pixels in figure~\ref{figure4a_4d}c. These highlight positions with a signal above a threshold value of 1125~counts. This threshold is given by adding the maximum error derived from the on- and off-resonance images to the background signal in the difference image. The background signal is determined by taking the average counts of the 25~pixels in the bottom left corner. Here, the HRTEM image (figure~\ref{figure4a_4d}d) clearly shows that no erbium is present. Overall, the structure of the Er map is in good agreement with the curved shape of the CNT bundle visible in the corresponding HRTEM image (figure~\ref{figure4a_4d}d).
\par
However, the regions with high intensity in the Er map appear about two times broader than the CNT bundle found in the HRTEM image. This can be explained by the beam spot size of 15\,nm.
\subsection{XMCD measurements of bulk samples}
It is of interest to know if the magnetic properties are affected by the encapsulation into CNTs. Characterizing the ground state magnetic properties of endofullerenes requires liquid helium temperatures and large external magnetic fields. These conditions cannot be met with the current STXM setup, and we have, therefore, performed an XMCD study of a macroscopic assembly of filled CNTs at a temperature of around 2\,K and magnetic fields up to 6.5\,T using an X\hbox{-}ray beam with a spot size of 1\,mm$^2$. The sample was prepared in a similar way as for the STXM measurements, but with a higher density of filled CNTs. As a reference, we started by performing an XMCD study of an Er$_3$N@C$_{80}$ bulk sample.
\par
\begin{figure}[h!]
	\begin{center}
		\includegraphics[width=16cm]{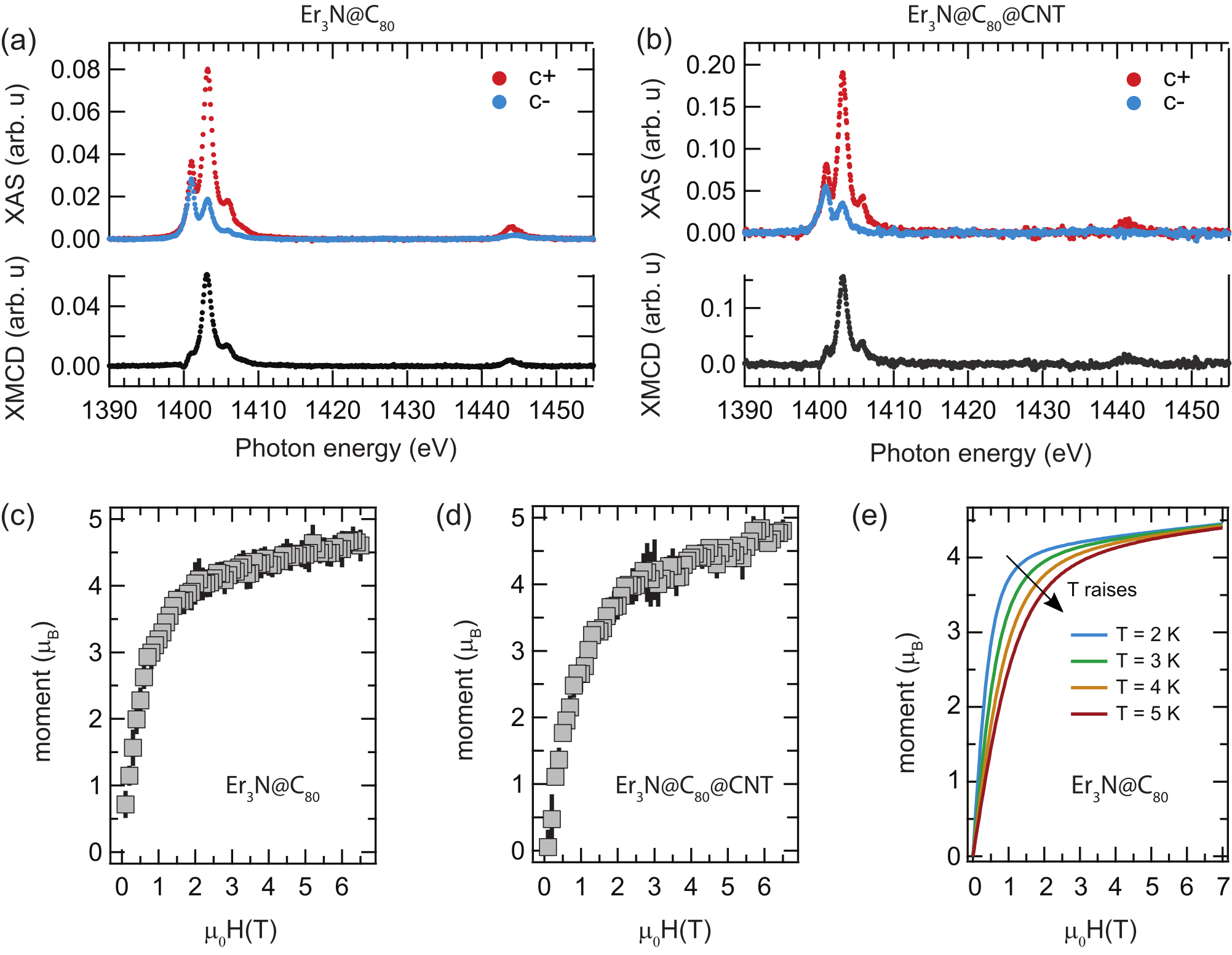}
		\caption{XAS and XMCD spectra from Er$_3$N@C$_{80}$ (a) and Er$_3$N@C$_{80}$@CNT (b), temperature set to 2\,K and magnetic field to 6.5\,T. Element specific magnetization curves recorded at the Er\,$M_5$\hbox{-}edge from Er$_3$N@C$_{80}$ (c) and Er$_3$N@C$_{80}$@CNT (d), temperature set to 2\,K. The magnetization curves correspond to the average of several independent measurements and the error bars shown in black are the standard deviation at each external magnetic field. The magnetization curves are scaled to the magnetic moments extracted from the sum rule analyses in table~\ref{table1}. (e) Simulated magnetization curves for non-interacting Er$^{3+}$ ions in the nitride clusterfullerene molecule.}
		\label{figure5a_5e}
	\end{center}
\end{figure}

Figure~\ref{figure5a_5e}a displays polarization dependent XAS and resulting XMCD spectra from the Er\,$M_{4,5}$\hbox{-}edges $(3d \rightarrow 4f)$, recorded from a bulk sample with the temperature set to 2\,K and with an external magnetic field of 6.5\,T applied along the X\hbox{-}ray beam. The spin $\langle S_z \rangle$ and orbital $\langle L_z \rangle$ angular momenta of the trivalent Er$^{3+}$ ions were extracted by applying a sum rule analysis \cite{Thole1992,Carra1993} and the results are presented in table~\ref{table1}. The number of holes was taken to be $n_h=3$ and $\langle T_z \rangle$ was calculated according to the literature \cite{Carra1993}. The $\langle L_z \rangle / \langle S_z \rangle$ ratio is lower than, but close to the value of $6/1.5=4$ expected from Hund's rules. The determined magnetic moment of $4.8\,\mu_B$ per ion corresponds to approximately half of the saturated magnetization of $9\,\mu_B$ expected for an isotropic paramagnet with $J=15/2$ and $g_J=6/5$. Additional deviations from an ideal paramagnet can be seen in the element specific magnetization curve displayed in figure~\ref{figure5a_5e}c. At a temperature of about 2\,K, the magnetization does not appear to reach saturation at a field up to 6.5\,T, suggesting that the ground state does not correspond to the maximum projection $m_J=\pm 15/2$.
\par
To obtain further information regarding the magnetic ground state we performed \textit{ab initio} calculations (for details see Experimental Section). In accordance with earlier point-charge calculations, \textit{ab initio} modelling shows that Er$^{3+}$ has easy-plane anisotropy \cite{Zhang2015}. The ground state is dominated by $m_J=\pm 1/2$ (66\%) with admixture of $m_J=\pm 3/2$ (28\%). In general, $J_z$ appears to be not a very good quantum number, especially for the low energy states. The crystal field splitting is predicted to be rather large, with the first excited state at 115\,cm$^{-1}$ and the highest excited state (which is in fact almost pure $m_J=\pm 15/2$ state) is found at 747\,cm$^{-1}$.
\par
Magnetization curves predicted for the powder sample are shown in figure~\ref{figure5a_5e}e. Possible exchange and dipolar interactions between Er$^{3+}$ ions within the Er$_3$N@C$_{80}$ molecule were not taken into account at this stage as they were shown to be quite small \cite{Tiwari2008}. For non-interacting Er$^{3+}$ ions in the nitride clusterfullerene molecule, we do not observe a saturation up to 7\,T, in agreement with experimental data. The magnetic moment of the Er ion is approaching $4.5\,\mu_B$ at 7\,T, which also agrees with experimental observations.
\par
The XMCD measurements were repeated on a macroscopic assembly of filled CNTs, and the results are presented in figure~\ref{figure5a_5e}b. Within the error bars, the sum rule analysis yields the same magnetic moments for the Er$_3$N@C$_{80}$ endofullerenes in the bulk phase and encapsulated in CNTs, see table~\ref{table1}. The element specific magnetization curve from the peapod sample is displayed in figure~\ref{figure5a_5e}d. It should be noted that the magnetization curves for the two samples cannot be directly compared since the temperature is not exactly known and the different mounting of the two samples could lead to different absolute temperatures. However, the magnetization curve has the same non-saturating behaviour as the bulk sample and simulations.
\begin{table}[h]
	\caption{\label{table1}Expectation values of spin $\langle S_z \rangle$ and orbital $\langle L_z \rangle$ angular momentum operators extracted from sum rule analyses of XMCD data (in units of $\hbar$).}
	\begin{center}
		\begin{tabular}{l c c c c c c c c} 
			\hline \hline
			Sample &$\langle S_z \rangle$& $\langle L_z \rangle$ & $\langle L_z \rangle / \langle S_z \rangle$ & $\langle L_z \rangle +2 \langle S_z \rangle$ \\ \hline
			Er$_3$N@C$_{80}$	& $0.9\pm0.1$ & $2.8\pm0.1$ & $3.1$ & 4.6 & \\
			Er$_3$N@C$_{80}$@CNT	& $0.9\pm0.2$ & $3.0\pm0.2$ & $3.3$ & 4.8 & \\
			\hline \hline
		\end{tabular}
	\end{center}
\end{table}
\par
XMCD measurements performed on Er$_3$N@C$_{80}$ and Er$_3$N@C$_{80}$@CNT have demonstrated that no significant change in the magnetic properties can be observed as a result of the encapsulation into CNTs. For both samples a magnetic moment of approximately half of that expected from an isotropic paramagnet is observed. A reduced effective magnetic moment of the Er$_3$N@C$_{80}$ endofullerene has previously been reported using SQUID magnetometry where the effect was explained by a quenching of the angular moment of the Er$^{3+}$ ions \cite{Tiwari2008,Smirnova2008}. No evidence of a quenched orbital moment could be observed in the present study. Instead, we attribute the reduced magnetic moment to the ligand field splitting and magnetic anisotropy of the Er$^{3+}$ ions.
\section{Conclusion}
In this paper, Er$_3$N@C$_{80}$ endofullerenes encapsulated into carbon nanotubes were investigated with HRTEM, STXM and XMCD. The successful filling was demonstrated with HRTEM measurements, while the X\hbox{-}ray measurements revealed the magnetic properties of the metallofullerene peapod assemblies.
\par
We demonstrated that the same peapod bundles can be localized and imaged with both electrons and X\hbox{-}rays, allowing us to combine the high spatial resolution of HRTEM with the local spectroscopy properties of STXM. Combining the two local techniques enables us to correlate an area with high Er concentration with a region occupied by a single peapod bundle, resulting in an element specific spatial resolution of about 30\,nm.
\par
XMCD measurements from an Er$_3$N@C$_{80}$ bulk sample and a macroscopic assembly of filled CNTs indicate that the magnetic properties of the endohedral Er$^{3+}$ ions are unchanged when the endofullerene molecules are encapsulated in CNTs.
\par
A local magnetic characterization of one-dimensional arrays of SMMs is technically challenging and would require XMCD measurements using X\hbox{-}ray microscopy techniques at low temperatures and with applied magnetic fields.
\par
We demonstrate that it is possible to record nanoscale XAS spectra from the $M_5$\hbox{-}edge of endohedral lanthanide ions in single nanoscale peapod bundles using X\hbox{-}ray microscopy. These measurements could be extended to spin-dependent absorption using circular polarized X\hbox{-}rays, thereby opening up the possibility of performing XMCD studies of low concentrations of lanthanide ions embedded in molecular nanostructures.
\acknowledgments
We thank the staff at the SLS for help: X\hbox{-}Treme beamline: C.~Piamonteze, PolLux beamline: B.~Sarafimov for technical support; The PolLux end station was financed by the Bundesministerium f\"ur Bildung und Forschung (BMBF) through contracts 05KS4WE1/6 and 05KS7WE1. We thank A.~N.~Khlobystov (Nottingham University) and T.~Chamberlain (University of Leeds) for fruitful discussions about the CO$_2$ filling and for information regarding the setup of the equipment. This work was financed by the Volkswagen Stiftung program ``Integration of Molecular Components in Functional Macroscopic Systems'' (Grant No. 86393 and 9200), the DFG SPP1601 and the Swedish Research Council (Grant No. 2015-00455). C.~M. acknowledges funding by the ``Nieders\"achsisches Vorab'' program of the Nieders\"achsisches Ministerium f\"ur Wissenschaft und Kultur. A.~A.~P. acknowledges funding by the European Research Council (ERC) under the European Union's Horizon 2020 research and innovation program (grant agreement No 648295 ``GraM3''). Computational resources were provided by the Center for Information Services and High Performance Computing (ZIH) in TU Dresden.

\bibliographystyle{apalike}
\end{document}